\documentclass[cits]{PoS}\usepackage{amsmath,cite}

\title{Strong Isospin Breaking in the Decay Constants of
Heavy--Light Mesons from Local-Duality QCD Sum Rules}
\ShortTitle{Strong Isospin Breaking in the Decay Constants of
Heavy--Light Mesons} 
\author{Wolfgang Lucha\\Institute for High Energy Physics, Austrian
Academy of Sciences, Nikolsdorfergasse 18,\\A-1050 Vienna,
Austria\\E-mail: \email{Wolfgang.Lucha@oeaw.ac.at}}
\author{\speaker{Dmitri Melikhov}\\Institute for High Energy
Physics, Austrian Academy of Sciences, Nikolsdorfergasse
18,\\A-1050 Vienna, Austria, and\\D.~V.~Skobeltsyn Institute of
Nuclear Physics, M.~V.~Lomonosov Moscow State University,\\119991,
Moscow, Russia, and\\Faculty of Physics, University of Vienna,
Boltzmanngasse 5, A-1090 Vienna, Austria\\E-mail:
\email{dmitri\_melikhov@gmx.de}}\author{Silvano Simula\\INFN,
Sezione di Roma Tre, Via della Vasca Navale 84, I-00146 Roma,
Italy\\E-mail: \email{simula@roma3.infn.it}}

\abstract{We explore, by a rather peculiar limit of QCD sum rules
subject to a Borel transformation (realized by allowing the
related Borel mass parameter to approach infinity), how the small
difference of the masses of up and down quark translates into the
leptonic decay constants of pseudoscalar or vector heavy--light
mesons. For the charmed and bottom mesons, we find that the decay
constants of their \emph{lowest-lying\/} charged and neutral
representatives should differ by an amount of the
order~of~$1\;\mbox{MeV}$.}

\FullConference{XIII Quark Confinement and the Hadron Spectrum ---
Confinement2018\\31 July -- 6 August 2018\\Maynooth University,
Ireland}

\begin{document}\section{Isospin Breaking in Weak Decay Constants
from Isospin-Violating Quark Masses}The comparatively small but
definitely nonzero difference of the masses of up and down quarks
$(m_d-m_u)(2\;\mbox{GeV})\approx2.5\;\mbox{MeV}$ \cite{PDG} causes
a mismatch of the leptonic decay constants of heavy--light mesons.
We discuss this phenomenon, for the case of the $D$, $D^\ast$,
$B$, and $B^\ast$ mesons, by considering a target-oriented
limiting case \cite{LMSLD1,LMSLD2,LMSLD3,LMSLD4} of the standard
formulation of the QCD sum-rule framework \cite{QSR}.

\section{QCD Sum-Rule Approach to Mesons, Borel Transformation,
Local-Duality Limit}\label{B}To begin with, let us rather
cursorily recall the QCD sum-rule description \cite{QSR} of hadron
systems formed by the strong interactions. Consider a heavy--light
meson bound state $(\bar q\,Q),$ in the following generically
called $H_q$, of a heavy quark $Q=c,b$ of mass $m_Q$, and a light
quark $q=u,d,s$ of mass $m_q.$ For any such meson, the basic
characteristics presently in the focus of our interest are its
mass, $M_{H_q},$ and its leptonic decay constant, $f_{H_q}$. Meson
properties of this kind are related, by QCD sum rules, to the
fundamental parameters of the underlying quantum field theory,
QCD,~\emph{viz.},~strong fine-structure coupling $\alpha_{\rm s}$,
quark masses, and vacuum condensates $\langle\bar
q\,q\rangle,\dots$, reflecting the nonperturbative side~of the
coin. Application of a Borel transformation from momentum to a new
variable dubbed the Borel parameter, $\tau,$ casts the QCD sum
rules in the focus of our efforts into a particularly convenient
form:\begin{align}f_{H_q}^2
\left(M_{H_q}^2\right)^N\exp(-M_{H_q}^2\,\tau)&=\hspace{-2.6ex}
\int\limits_{(m_Q+m_q)^2}^{s^{(N)}_{\rm
eff}(\tau,m_Q,m_q,\alpha_{\rm s})}\hspace{-2.6ex}{\rm d}s
\exp(-s\,\tau)\,s^N\,\rho(s,m_Q,m_q,\alpha_{\rm s}\,|\,m_{\rm
sea})\nonumber\\&+\Pi^{(N)}_{\rm power}(\tau,m_Q,m_q,\alpha_{\rm
s},\langle\bar q\,q\rangle,\dots)\ .\label{s}\end{align}The
right-hand side of this relation involves a variety of ingredients
governed, in principle, by~QCD:\begin{itemize}\item The spectral
density $\rho(s,m_Q,m_q,\alpha_{\rm s}\,|\,m_{\rm sea})$
encompasses all purely perturbative contributions, derivable by
series expansion in powers of $\alpha_{\rm s}(\mu)$ (depending on
the renormalization scale~$\mu$),\begin{align}
\rho(s,m_Q,m_q,\alpha_{\rm s}\,|\,m_{\rm sea})&=\rho_0(s,m_Q,m_q)
+\frac{\alpha_{\rm s}(\mu)}{\pi}\,\rho_1(s,m_Q,m_q,\mu)\nonumber\\
&+\frac{\alpha_{\rm s}^2(\mu)}{\pi^2}\,
\rho_2(s,m_Q,m_q,\mu\,|\,m_{\rm sea})+O(\alpha_{\rm s}^3)\
.\label{SD}\end{align}Starting at order $\alpha_{\rm s}^2$, it
depends also on the masses of all sea quarks, collectively
labelled~$m_{\rm sea}$.\item The power corrections $\Pi^{(N)}_{\rm
power}(\tau,m_Q,m_q,\alpha_{\rm s},\langle\bar q\,q\rangle,\dots)$
describe nonperturbative contributions by means of vacuum
condensates of products of quark and gluon fields, brought into
the game by Wilson's operator product expansion \cite{KGW},
accompanied by powers of $\tau$, whence that~name.\item The
effective threshold $s^{(N)}_{\rm eff}(\tau,m_Q,m_q,\alpha_{\rm
s})$ delimits, as its lower boundary, the range of values of our
integration variable $s$ where, according to the postulate of
global quark--hadron duality, mutual cancellation of
perturbative-QCD and hadron contributions takes place. This
quantity definitely depends, in general, also on the Borel
parameter $\tau$ \cite{LMST1,LMST2,LMST3,LMST4,LMST5}. Taking into
account its $\tau$ dependence increases considerably the accuracy
of one's QCD sum-rule predictions \cite{LMSA,LMSA1,LMSA2,LMSA3}.
\item The non-negative integer $N=0,1,2,\dots$ can be used to
optimize the study: it decides upon the share of power corrections
and effective threshold in the entire nonperturbative
contributions.\end{itemize}\pagebreak

The Borel parameter $\tau$ is a free variable yet to be determined
or chosen. Within the formulation of conventional Borelized QCD
sum rules, the meaningful region of $\tau$ gets constrained, from
below, by insisting on a utilizably large ground-state
contribution and, from above, by requiring reasonably small
power-correction contributions. In contrast, the local-duality
limit is defined by letting $\tau\to0$.

If, for a convenient choice of $N$, all power corrections vanish
in the local-duality limit, that~is,~if
$$\lim_{\tau\to0}\Pi^{(\widehat N)}_{\rm
power}(\tau,m_Q,m_q,\alpha_{\rm s},\langle\bar
q\,q\rangle,\dots)=0\qquad\mbox{for some $\widehat N$}\ ,$$the
generic Borelized QCD sum rule (\ref{s}) reduces to a spectral
integral over its perturbative density:\begin{align}f_{H_q}^2
\left(M_{H_q}^2\right)^{\widehat
N}=\hspace{-2.58ex}\int\limits_{(m_Q+m_q)^2}^{s^{(\widehat
N)}_{\rm eff}(m_Q,m_q,\alpha_{\rm s})}\hspace{-2.58ex}{\rm
d}s\,s^{\widehat N}\,\rho(s,m_Q,m_q,\alpha_{\rm s}\,|\,m_{\rm
sea})\ ,&\label{0}\\s^{(\widehat N)}_{\rm eff}(m_Q,m_q,\alpha_{\rm
s})\equiv s^{(\widehat N)}_{\rm eff}(0,m_Q,m_q,\alpha_{\rm s})\
.&\nonumber\end{align}In this case, the effective threshold
$s^{(\widehat N)}_{\rm eff}(m_Q,m_q,\alpha_{\rm s})$ must take
care of all the nonperturbative effects.

\section{Illustrative Special Case: Pseudoscalar and Vector-Meson
Weak Decay Constants}
\begin{figure}[h]\centering
\includegraphics[scale=.67548,clip]{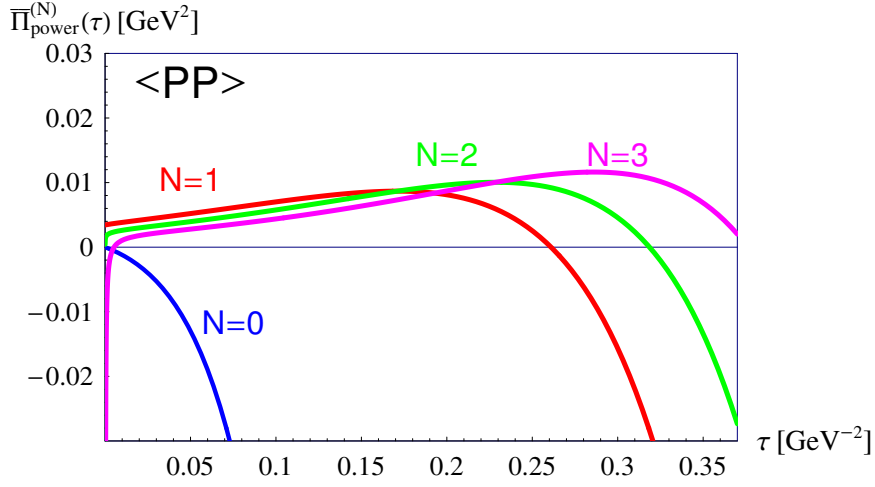}\\(a)\\[1ex]
\includegraphics[scale=.67548,clip]{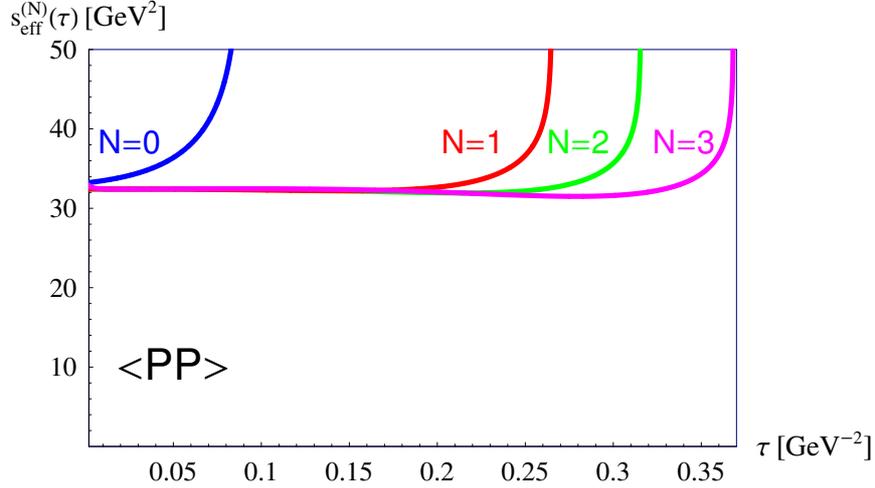}\\(b)
\caption{Pseudoscalar mesons: $\tau$ dependence of (a) power
corrections (\protect\ref{pc}) and (b) effective
threshold~$s^{(N)}_{\rm eff}$.}\label{P}\end{figure}

\begin{figure}[h]\centering
\includegraphics[scale=.67548,clip]{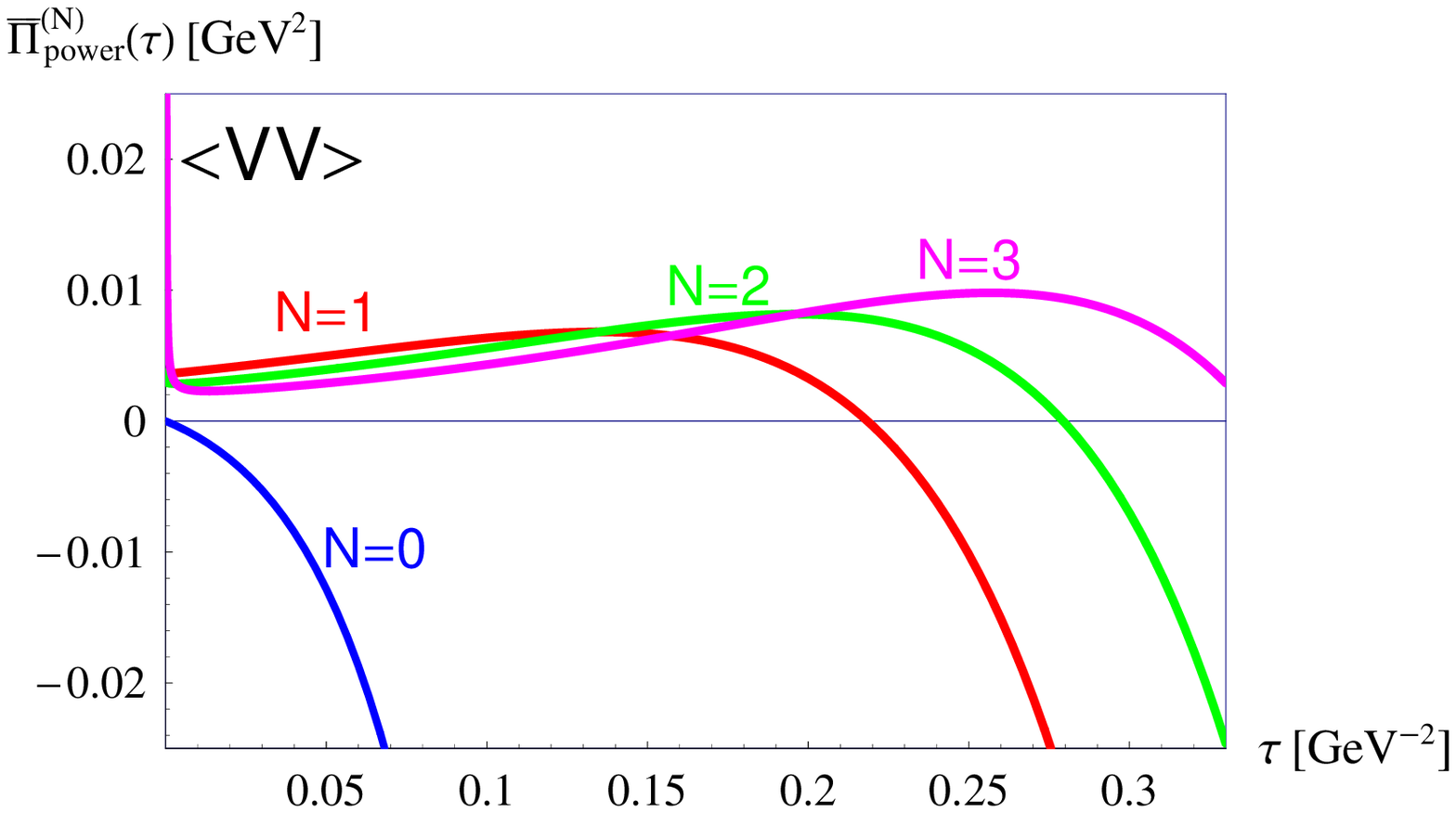}\\(a)\\[1ex]
\includegraphics[scale=.67548,clip]{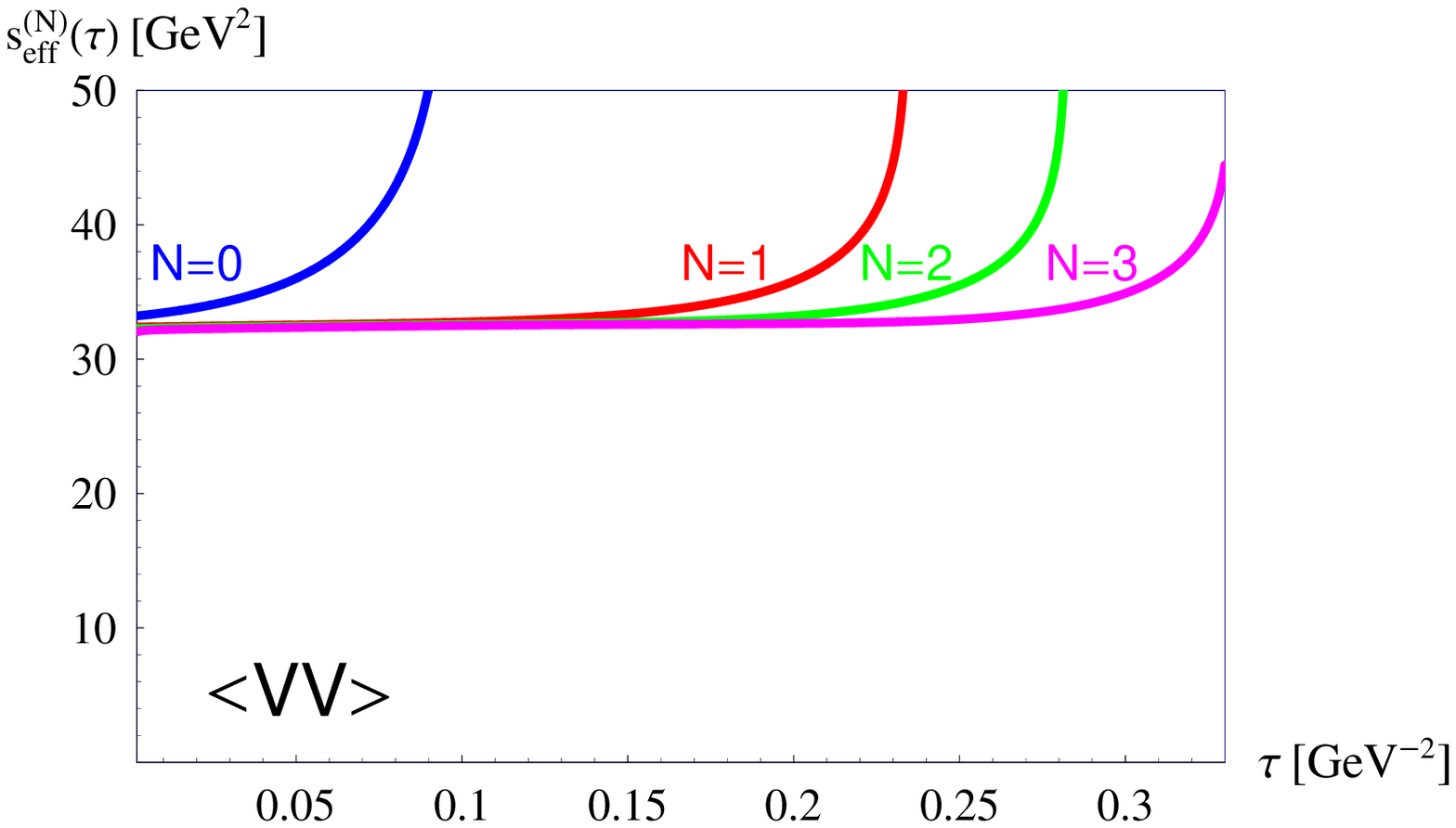}\\(b)
\caption{Vector mesons: $\tau$ dependence of (a) the power
corrections (\protect\ref{pc}) and (b) the effective
threshold~$s^{(N)}_{\rm eff}$.}\label{V}\end{figure}

Let us now boil down the general QCD sum-rule
approach to mesons briefly sketched in Sect.~\ref{B} to our actual
targets, heavy-light pseudoscalar and vector mesons, \emph{i.e.},
$H_q=P_q,V_q$, of momentum~$p$. Their weak decay constants,
$f_{P_q,V_q}$, result from matrix elements of the relevant
axial-vector or vector heavy--light quark current: for a
pseudoscalar meson $P_q$, its leptonic decay constant $f_{P_q}$ is
defined~by$$\langle0|\,\bar
q(0)\,\gamma_\mu\,\gamma_5\,Q(0)\,|P_q(p)\rangle={\rm i}\,f_{P_q}
\,p_\mu\ ;$$for a vector meson $V_q$, with polarization vector
$\varepsilon_\mu(p)$, its weak decay constant $f_{V_q}$ can be
found~from$$\langle0|\,\bar q(0)\,\gamma_\mu\,Q(0)\,
|V_q(p)\rangle=f_{V_q}\,M_{V_q}\,\varepsilon_\mu(p)\ .$$For all
such mesons, the spectral densities are available up to order
$O(\alpha_{\rm s}\,m_q)$ and $O(\alpha_{\rm s}^2\,m_q^0)$
\cite{d1,d2,d3,d4}. The applicability of the local-duality limit
is endangered by the potential existence of terms of order
\begin{equation}\tau^{N-2}\log(\tau)\label{t}\end{equation}in the
power corrections. Accordingly, let us inspect, in Figs.~\ref{P}
and \ref{V}, for the cases $N=0,1,2,3$, the $\tau$ dependence of
the power corrections and spectral densities, the former being
suitably redefined~by
\begin{equation}\overline{\Pi}^{(N)}_{\rm power}(\tau)\equiv
\Pi^{(N)}_{\rm power}(\tau,\dots)\,\frac{\exp(M_{H_q}^2\,\tau)}
{\left(M_{H_q}^2\right)^N}\ .\label{pc}\end{equation}In accordance
with the behaviour expectable from expression (\ref{t}), all the
power corrections vanish for $N=0$, approach a finite value for
$N=1$, and develop a singularity for $N=2,3$ in the
limit~$\tau\to0$:\begin{align*}\lim_{\tau\to0}\Pi^{(0)}_{\rm
power}(\tau,m_Q,m_q,\alpha_{\rm s},\langle\bar
q\,q\rangle,\dots)&=0\ ,\\\lim_{\tau\to0}|\Pi^{(1)}_{\rm
power}(\tau,m_Q,m_q,\alpha_{\rm s},\langle\bar
q\,q\rangle,\dots)|&<\infty\ ,\\\lim_{\tau\to0}|\Pi^{(2,3)}_{\rm
power}(\tau,m_Q,m_q,\alpha_{\rm s},\langle\bar
q\,q\rangle,\dots)|&=\infty\ .\end{align*}The local-duality limit
of $\widehat N=0$ QCD sum rules for pseudoscalar and vector mesons
is well-defined.\pagebreak

Choosing $\widehat N=0$ simplifies the QCD sum rule (\ref{0}) to a
relation for only the decay constant~$f_{H_q}$:\begin{equation}
f_{H_q}^2=\hspace{-2.3ex}\int\limits_{(m_Q+m_q)^2}^{s_{\rm
eff}(m_Q,m_q,\alpha_{\rm s})}\hspace{-2.3ex}{\rm d}s\,
\rho_{H_q}(s,m_Q,m_q,\alpha_{\rm s}\,|\,m_{\rm sea})\equiv
\digamma_{\!\!H_q}(s_{\rm eff}(m_q),m_q\,|\,m_{\rm sea})\ ,\qquad
H_q=P_q,V_q\ .\label{f}\end{equation}We denote the function
encoding all light-quark-mass dependence by the Greek letter
$\digamma\!$ (digamma).

\section{Strong Isospin Breaking in Charmed- and Bottom-Meson Weak
Decay Constants}Heavy--light-meson decay-constant studies
\cite{LMSC1,LMSC2,LMSC3,LMSC4,LMSC5} relying on the conventional
QCD sum-rule framework with Borel-parameter-upgraded effective
threshold draw the behaviour of the latter~from knowledge about
meson \emph{masses\/} in our focus. In the local-duality approach,
upon noting that \cite{LMSLD1,LMSLD3,LMSLD4}

\pagebreak
\begin{figure}[hbt]\centering
\includegraphics[scale=.52571,clip]{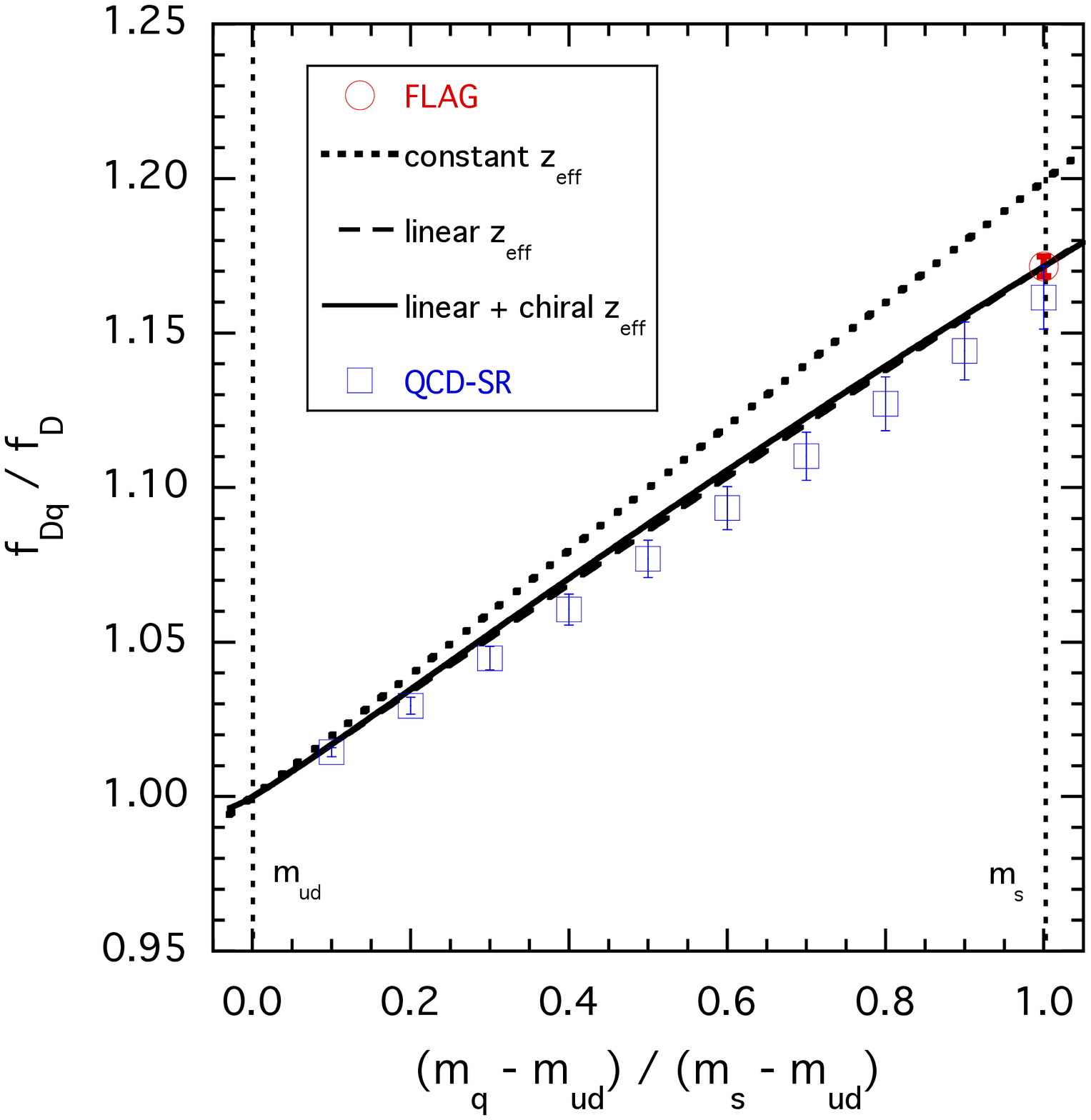}\\[2ex]
\includegraphics[scale=.52571,clip]{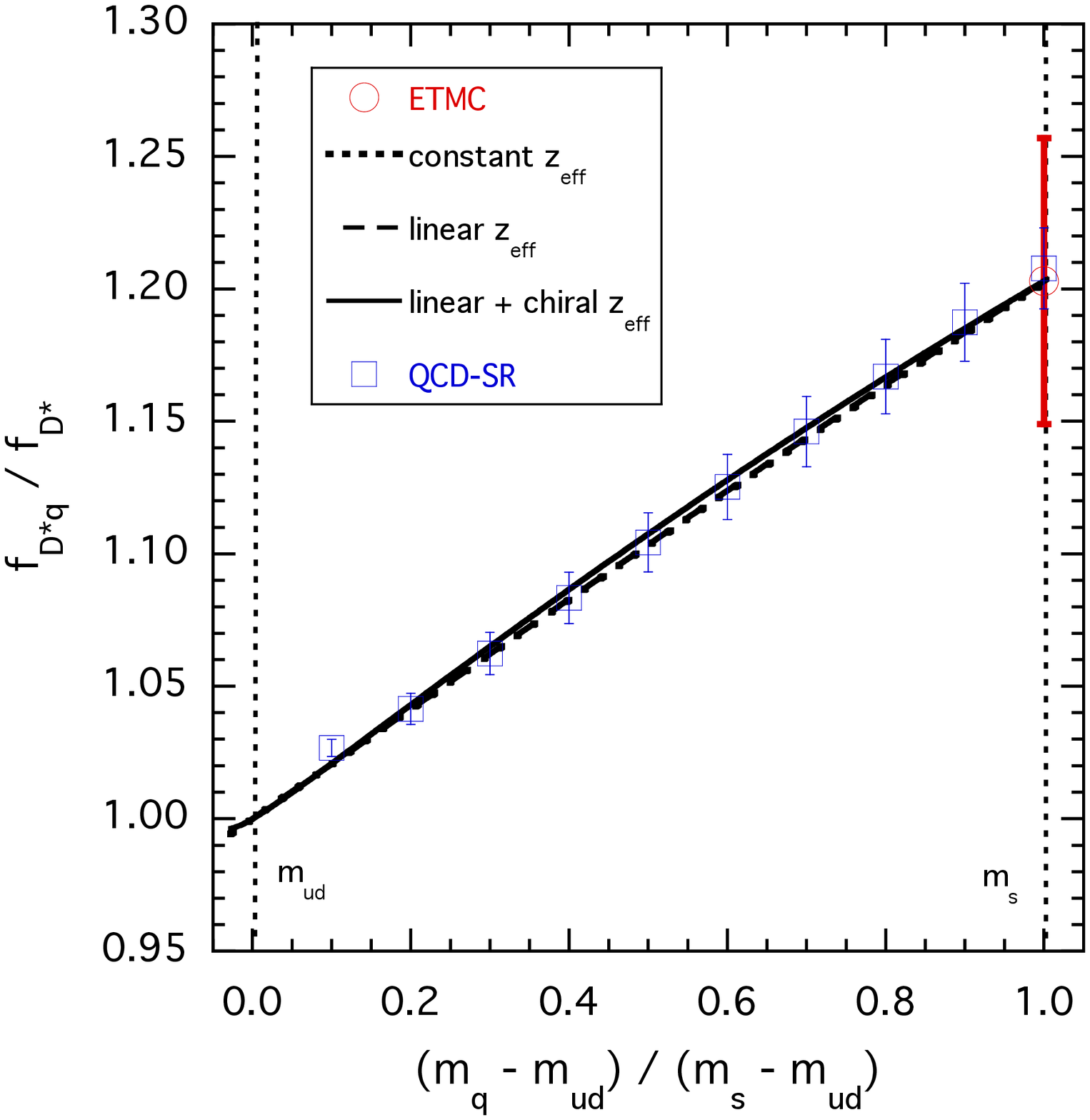}
\caption{Behaviour of the decay constants $f_{D^{(\ast)}_q}\equiv
f_{D^{(\ast)}_q}(m_q)$ of charmed pseudoscalar and vector mesons,
in units of $f_{D^{(\ast)}}\equiv f_{D^{(\ast)}_q}(m_{ud})$, for
the light-quark mass $m_q$ varying in the interval $(0,m_s)$,
resulting from the~three ansatzes considered by Ref.~\cite{LMSLD1}
(dubbed ``constant'', ``linear'' and ``linear + chiral'') for our
effective-threshold function $z_{\rm eff}$ defined in
Eq.~(\protect\ref{z}) \cite{LMSLD1,LMSLD2,LMSLD3,LMSLD4}, compared
with the results of a conventional QCD sum-rule~study
\cite{LMSIB}.}\label{D:f/f}\end{figure}
\begin{figure}[hbt]\centering
\includegraphics[scale=.52571,clip]{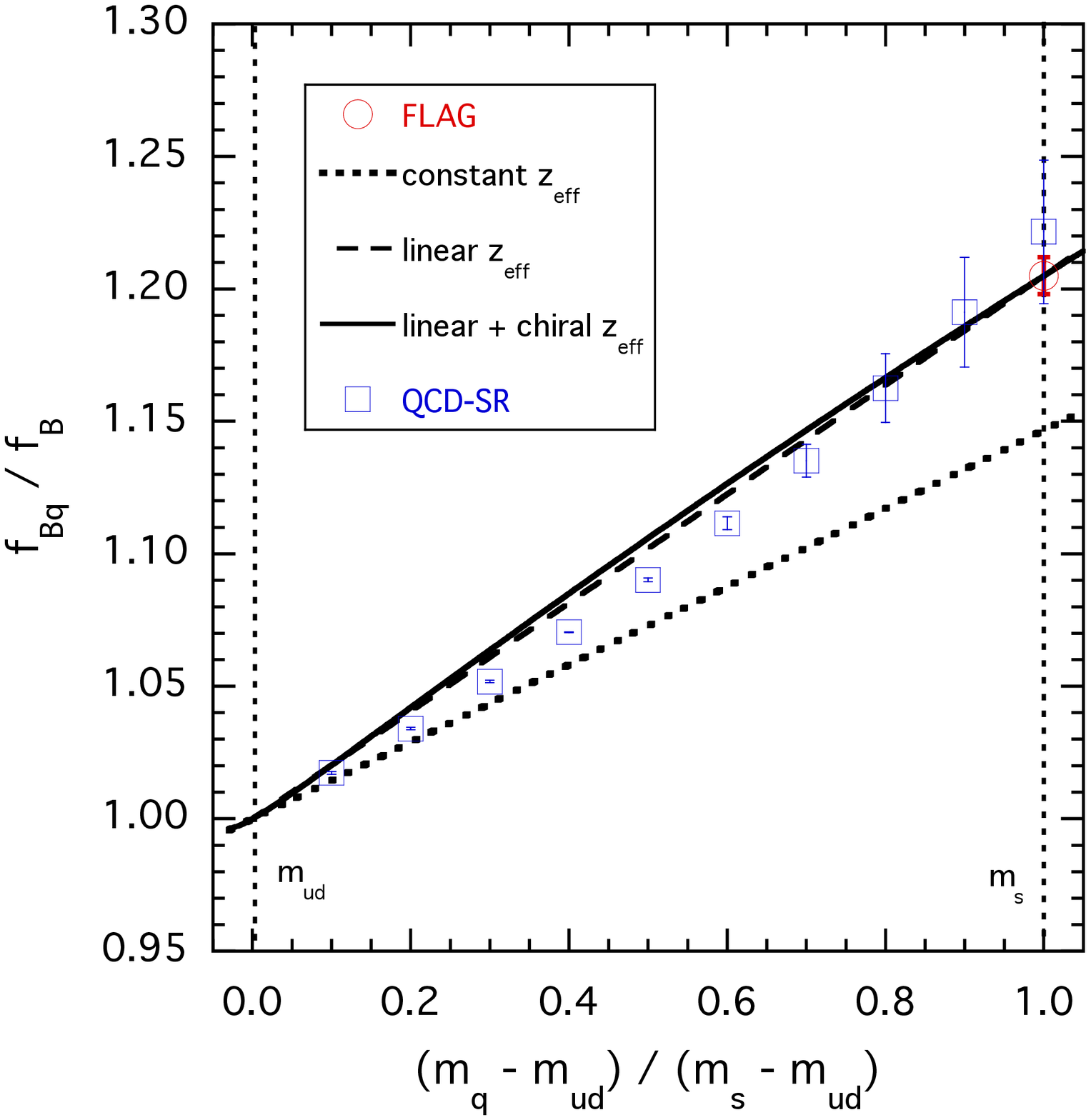}\\[2ex]
\includegraphics[scale=.52571,clip]{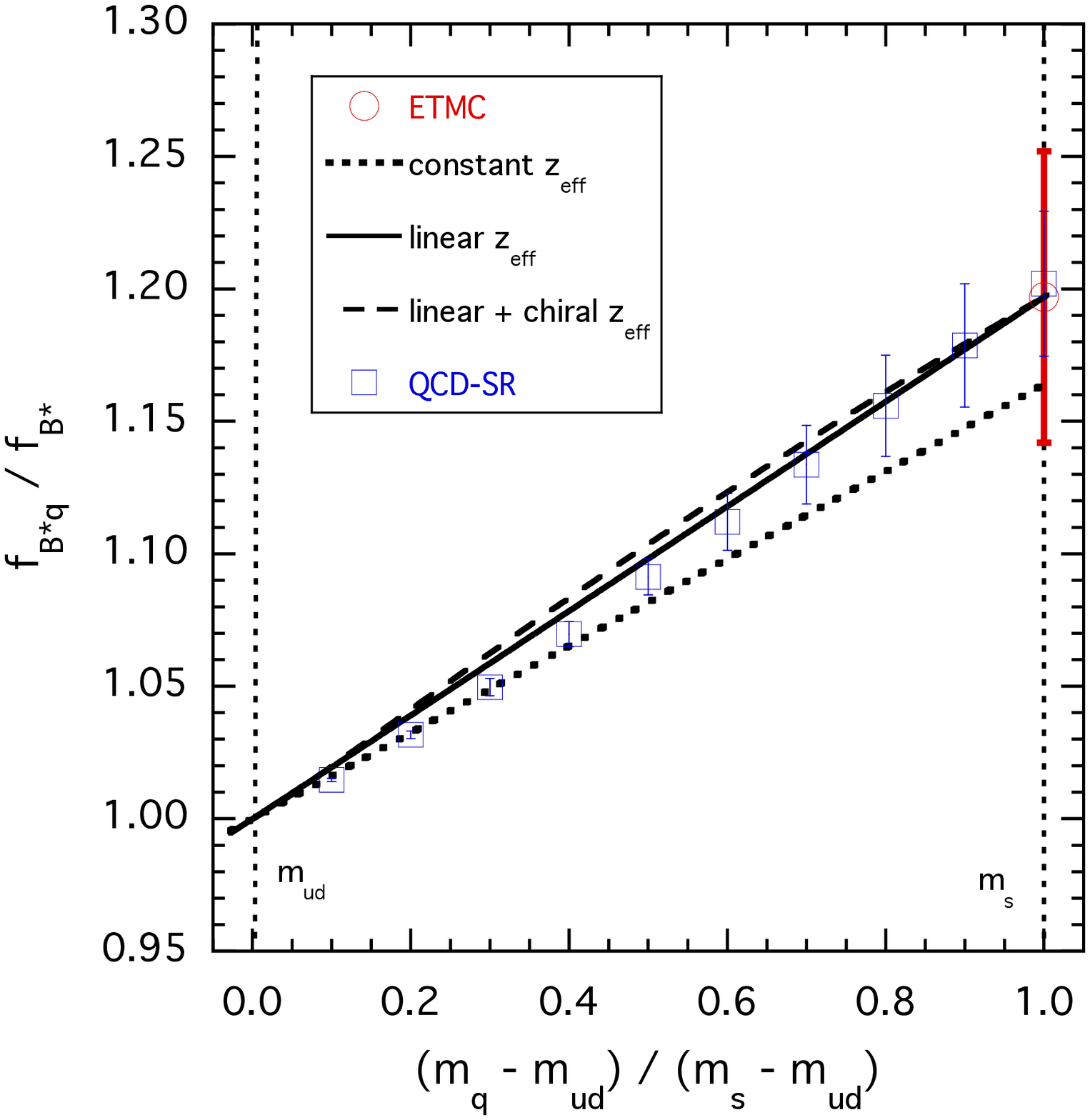}
\caption{Behaviour of the decay constants $f_{B^{(\ast)}_q}\equiv
f_{B^{(\ast)}_q}(m_q)$ of both pseudoscalar and vector bottom
mesons, in units of $f_{B^{(\ast)}}\equiv
f_{B^{(\ast)}_q}(m_{ud})$, for the light-quark mass $m_q$ varying
in the interval $(0,m_s)$, arising from the three ansatzes
considered by Ref.~\cite{LMSLD1} (dubbed ``constant'', ``linear''
and ``linear + chiral'') for our effective-threshold function
$z_{\rm eff}$ defined in Eq.~(\protect\ref{z})
\cite{LMSLD1,LMSLD2,LMSLD3,LMSLD4}, compared with the results of a
conventional QCD sum-rule~study \cite{LMSIB}.}\label{B:f/f}
\end{figure}


\noindent in the difference of heavy--light decay constants the
dependence on the strange sea quarks drops~out,\begin{align}&
\digamma_{\!\!H_d}(s_{\rm eff}(m_d),m_d\,|\,m_{\rm sea})-
\digamma_{\!\!H_u}(s_{\rm eff}(m_u),m_u\,|\,m_{\rm sea})\nonumber
\\&=
\digamma_{\!\!H_d}(s_{\rm eff}(m_d),m_d\,|\,m_{\rm sea}=0)-
\digamma_{\!\!H_u}(s_{\rm eff}(m_u),m_u\,|\,m_{\rm sea}=0)
+O\!\!\left(\frac{\alpha_{\rm s}^2}{\pi^2}\,(m_d-m_u)\right)
,\label{d}\end{align}we adjust the function $f_H(m_q)$ of a
continuously varying $m_q$, normalized to its average
$f_H(m_{ud})$~with $m_{ud}\equiv\frac{1}{2}\,(m_u+m_d)$, for
several ansatzes of increasing degree of sophistication for the
problem's~core\begin{equation}z_{\rm eff}\equiv\sqrt{s_{\rm
eff}}-m_Q-m_q\ ,\label{z}\end{equation}to related lattice-QCD
results $f_H(m_{ud})$ and $f_H(m_s)$ at a physical quark mass
\cite{L1,L2}. Figures \ref{D:f/f} and \ref{B:f/f} then yield, in
nearly perfect agreement with the conventional findings
\cite{LMSIB}, the $f_{H_q}$ differences
\cite{LMSLD1,LMSLD2,LMSLD3,LMSLD4}

\clearpage

\begin{align*}f_{D^\pm}-f_{D^0}&=(0.96\pm0.09)\;\mbox{MeV}\ ,&
f_{D^{*\pm}}-f_{D^{*0}}&=(1.18\pm0.35)\;\mbox{MeV}\ ,\\
f_{B^0}-f_{B^\pm}&=(1.01\pm0.10)\;\mbox{MeV}\ ,&
f_{B^{*0}}-f_{B^{*\pm}}&=(0.89\pm0.30)\;\mbox{MeV}\ .\end{align*}

\vspace{3.3873ex}\noindent{\bf Acknowledgement}. D.~M.\ is
supported by the Austrian Science Fund (FWF), project P29028-N27.


\begin{thebibliography}{99}
\bibitem{PDG}Particle Data Group (M.~Tanabashi \emph{et al.}),
Phys.~Rev.~D {\bf 98} (2018) 030001.
\bibitem{LMSLD1}W.~Lucha, D.~Melikhov, and S.~Simula,
Eur.~Phys.~J.~C {\bf 78} (2018) 168, arXiv:1702.07537 [hep-ph].
\bibitem{LMSLD2}W.~Lucha, D.~Melikhov, and S.~Simula,
PoS (EPS-HEP 2017) 669, arXiv:1709.02131 [hep-ph].
\bibitem{LMSLD3}W.~Lucha, D.~Melikhov, and S.~Simula,
PoS (Hadron2017) 175, arXiv:1711.07899 [hep-ph].
\bibitem{LMSLD4}W.~Lucha, D.~Melikhov, and S.~Simula, preprint
HEPHY-PUB 1006/18 (2018), arXiv:1808.07684 [hep-ph].
\bibitem{QSR}M.~A.~Shifman, A.~I.~Vainshtein, and V.~I.~Zakharov,
Nucl.~Phys.~B {\bf 147} (1979) 385.
\bibitem{KGW}K.~G.~Wilson, Phys.~Rev.~{\bf 179} (1969) 1499.
\bibitem{LMST1}W.~Lucha, D.~Melikhov, and S.~Simula, Phys.~Rev.~D
{\bf 79} (2009) 096011, arXiv:0902.4202 [hep-ph].
\bibitem{LMST2}W.~Lucha, D.~Melikhov, and S.~Simula, J.~Phys.~G
{\bf 37} (2010) 035003, arXiv:0905.0963 [hep-ph].
\bibitem{LMST3}W.~Lucha, D.~Melikhov, H.~Sazdjian, and S.~Simula,
Phys.~Rev.~D {\bf 80} (2009) 114028, arXiv:0910.3164 [hep-ph].
\bibitem{LMST4}W.~Lucha, D.~Melikhov, and S.~Simula, Phys.~Lett.~B
{\bf 687} (2010) 48, arXiv:0912.5017 [hep-ph].
\bibitem{LMST5}W.~Lucha, D.~Melikhov, and S.~Simula,
Phys.~Atom.~Nucl.~{\bf 73} (2010) 1770, arXiv:1003.1463 [hep-ph].
\bibitem{LMSA}W.~Lucha, D.~Melikhov, and S.~Simula, Phys.~Rev.~D
{\bf 76} (2007) 036002, arXiv:0705.0470 [hep-ph].
\bibitem{LMSA1}W.~Lucha, D.~Melikhov, and S.~Simula, Phys.~Lett.~B
{\bf 657} (2007) 148, arXiv:0709.1584 [hep-ph].
\bibitem{LMSA2}W.~Lucha, D.~Melikhov, and S.~Simula, Phys.~Lett.~B
{\bf 671} (2009) 445, arXiv:0810.1920 [hep-ph].
\bibitem{LMSA3}D.~Melikhov, Phys.~Lett.~B {\bf 671} (2009) 450,
arXiv:0810.4497 [hep-ph].
\bibitem{d1}K.~G.~Chetyrkin and M.~Steinhauser, Phys.~Lett.~B {\bf
502} (2001) 104, arXiv:hep-ph/0012002.
\bibitem{d2}K.~G.~Chetyrkin and M.~Steinhauser, Eur.~Phys.~J.~C
{\bf 21} (2001) 319, arXiv:hep-ph/0108017.
\bibitem{d3}M.~Jamin and B.~O.~Lange, Phys.~Rev.~D {\bf 65} (2002)
056005, arXiv:hep-ph/0108135.
\bibitem{d4}P.~Gelhausen, A.~Khodjamirian, A.~A.~Pivovarov, and
D.~Rosenthal, Phys.~Rev.~D {\bf 88} (2013) 014015, arXiv:1305.5432
[hep-ph]; Phys.~Rev.~D {\bf 89} (2014) 099901(E); {\bf 91} (2015)
099901(E).
\bibitem{LMSC1}W.~Lucha, D.~Melikhov, and S.~Simula, J.~Phys.~G
{\bf 38} (2011) 105002, arXiv:1008.2698 [hep-ph].
\bibitem{LMSC2}W.~Lucha, D.~Melikhov, and S.~Simula, Phys.~Lett.~B
{\bf 701} (2011) 82, arXiv:1101.5986 [hep-ph].
\bibitem{LMSC3}W.~Lucha, D.~Melikhov, and S.~Simula, Phys.~Rev.~D
{\bf 88} (2013) 056011, arXiv:1305.7099 [hep-ph].
\bibitem{LMSC4}W.~Lucha, D.~Melikhov, and S.~Simula, Phys.~Lett.~B
{\bf 735} (2014) 12, arXiv:1404.0293 [hep-ph].
\bibitem{LMSC5}W.~Lucha, D.~Melikhov, and S.~Simula, Phys.~Rev.~D
{\bf 91} (2015) 116009, arXiv:1504.03017 [hep-ph].
\bibitem{L1}FLAG Working Group (S.~Aoki \emph{et al.}),
Eur.~Phys.~J.~C {\bf 77} (2017) 112, arXiv:1607.00299 [hep-lat].
\bibitem{L2}ETM Collaboration (V.~Lubicz \emph{et al.}),
Phys.~Rev.~D {\bf 96} (2017) 034524, arXiv:1707.04529 [hep-lat].
\bibitem{LMSIB}W.~Lucha, D.~Melikhov, and S.~Simula, Phys.~Lett.~B
{\bf 765} (2017) 365, arXiv:1609.05050 [hep-ph].\pagebreak
\end{thebibliography}
\end{document}